\documentclass[superscriptaddress,pre,aps,twocolumn,showpacs,floatfix]{revtex4}

\usepackage[dvips]{graphicx}
\usepackage{amsmath}
\usepackage{amssymb}
\usepackage[nice]{nicefrac}
\usepackage{color}

\newcommand{\pin}{p_{\mathrm{new}}}
\newcommand{\pr}{p_{\mathrm{repeat}}}
\newcommand{\pcc}{p_{\mathrm{cc}}}
\newcommand{\rmin}{r_{\mathrm{min}}}

\begin{document}
\bibliographystyle{prsty}
\title{Information spreading and development of cultural centers}

\author{Bart{\l}omiej Dybiec}
\email{bartek@th.if.uj.edu.pl}
\affiliation{Center for Models of Life, Niels Bohr Institute, University of Copenhagen, Blegdamsvej 17, 2100 Copenhagen \O, Denmark}
\affiliation{Marian Smoluchowski Institute of Physics, and Mark Kac Center for Complex Systems Research, Jagiellonian University, ul. Reymonta 4, 30--059 Krak\'ow, Poland }

\author{Namiko Mitarai}
\email{mitarai@nbi.dk}
\affiliation{Center for Models of Life, Niels Bohr Institute, University of Copenhagen, Blegdamsvej 17, 2100 Copenhagen \O, Denmark}

\author{Kim Sneppen}
\email{sneppen@nbi.dk}
\affiliation{Center for Models of Life, Niels Bohr Institute, University of Copenhagen, Blegdamsvej 17, 2100 Copenhagen \O, Denmark}

\date{\today}
\begin{abstract}
The historical interplay between societies are governed by many factors,
including in particular spreading of languages, religion and other symbolic traits.
Cultural development, in turn, is coupled
to emergence and maintenance of information spreading.
Strong centralized cultures exist thanks to attention
from their members, which faithfulness in turn relies on supply of information.
Here, we discuss a culture evolution model on a planar geometry
that takes into account aspects of the feedback
between information spreading and its maintenance.
Features of model are highlighted by comparing it to
cultural spreading in ancient and medieval Europe,
where it in particular suggests that long lived centers should be located
in geographically remote regions.
\end{abstract}

\maketitle

\section{Introduction\label{sec:introduction}}

The expansion and decline of social structures depend on
information spreading in form of languages, religion,
or other cultural inventions \cite{ackland2007}.
In the recent years many mathematical models have been proposed for
social interactions and dynamics, trying to understand social structures.
However, a main feature of most models of culture dissemination is an
adaptation toward local or global consensus
\cite{axelrod1997,CastellanoReview2009}, an equilibration
which is also found in voter models \cite{holley1975}, social impact
theory \cite{lewenstein1992}, majority rules \cite{galam2002}, the Sznajds model
\cite{sznajd-weron2000}, the Deffuant model \cite{deffuant2000}
and the bounded confidence models \cite{hegselmann2002}.

The multitude of models that emphasize consensus-dynamics
contrast a reality where consensus is often broken by emergence
of new cultures, languages or opinions.
One driving force for heterogeneity is the need for attention,
where individuals not only aim at mutual understanding,
but at the same time also fight for individual attention.
This attention battle is more than
 random fluctuations of agents \cite{pineda2011}
 or rejection of other opinions \cite{huet2008}:
The battle may, for example, involve positive
feedback mechanisms as suggested by \cite{rosvall2009}.
In this paper,  we take the possibility for a new culture to emerge
into account, in addition to the local alignment rules.
Lacking a simple realistic mechanism for creation of new cultures,
we here simply parametrize this ``emergence'' in terms of
a rate $\pin$ for initiation of new cultures.

\begin{figure}[!h]
\centering
\includegraphics[width=0.95\columnwidth]{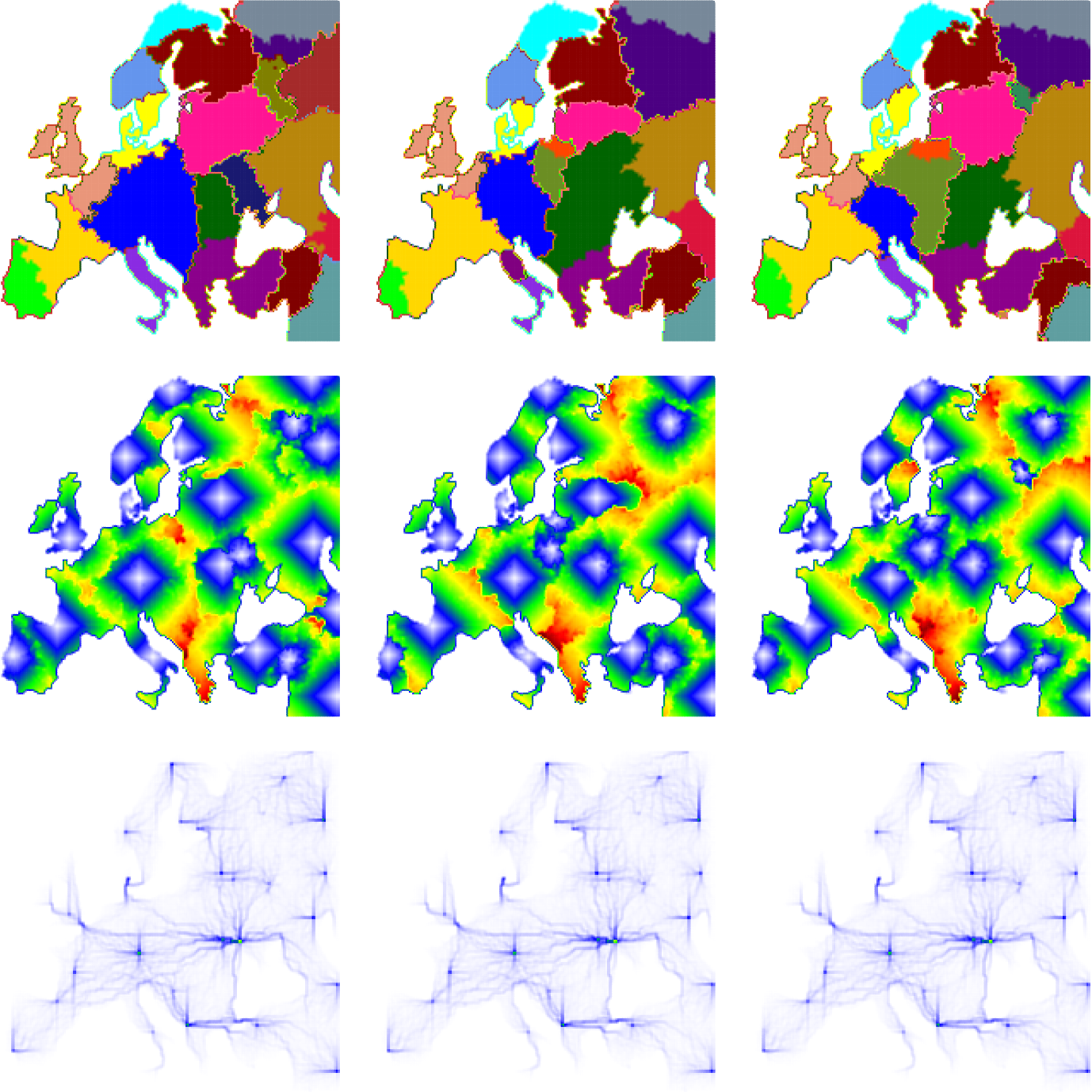}
\caption{(Color on-line) System snapshots at various times
$t=\{47, 48, 49\} \times 10^3$ (from left to right),
i.e. consecutive columns present subsequent states of the system separated by
1~000 Monte Carlo steps.
Different panels demonstrates: spatial structure of cultures (top panel),
distance to the cultural center (middle panel) and information pathways (bottom panel).
Cultural centers are located on the ``white-blue stars'' in the middle panels.
Parameters: $p=0.01, \pin=0.004$ and $\pr=0.015$.}
 \label{fig:timeseries}
\end{figure}

Another common theme of models dealing with the spread of information
is that two different pieces of information are treated on an
equal footing. This is in general an incorrect assumption, as the
importance of two bits of information in general is asymmetric.
One sorting principle is to use the information age as a sorting criteria,
reflecting the fact that the value of information typically decays with time
\cite{stiglitz1981credit}.
Previous studies \cite{rosvall2006self,lizana2010time} demonstrated that
such a sorting principle has major consequences for the spatio-temporal
dynamics of information.  Importantly,
 newer information overriding older information  has
 been observed in spreading of linguistic features
 \cite{Yanagita,ramsey1982language}.
 Furthermore,
a simple model of diffusion of information from
a cultural stronghold with age sorting is shown to be compatible with the observed
pattern of word distribution in Japan \cite{lizana2011}.

When individuals sort information based on its age,
an existing cultural center will continuously need to generate new
information to maintain their sphere of influence.
Accordingly, we here characterize the strength
of a cultural center by the rate with which it is
able to generate fashions, $\pr$.
We will take this rate as a characteristic of a cultural center,
and it keeps generating new fashions  until the center is eliminated by information generated
from competing cultural centers.

A main feature of the famous Axelrods model
\cite{axelrod1997,CastellanoReview2009}
for social alignment is conservativeness
in communication, implemented by having individuals
with many types of opinions and a preference for communication
between individuals that share many traits.
This preference  makes people more open
for communication towards sources where they earlier
obtained information.
We here parametrize such conservativeness into a single parameter, $p$,
that counts the chance that a given ``site'' or ``agent'' changes
who he prefers to obtain information from.

Overall our model aim to discuss
the information flow association to emerging and collapsing cultural centers,
each influencing their surroundings by an ongoing generation of announcements
that maintain their sphere of influence.
The details of the model are presented in the next section.

\section{Model \label{sec:model}}

Our model considers many rumors / fashions / viewpoints / stories / ideas
(denoted fashions in the following)
competing on a two dimensional square lattice of sites
which at any given moment can be occupied by one fashion only.
We imagine each site as an agent, which in fact could be
a whole group of people that by definition share the same taste.
Each site listens to their immediate four neighboring sites
with a history dependent frequency.
When communicating, they accept fashions
only when they are newer than
the current local fashion.
History, or conservativeness, is quantified in terms of
a preference in listening towards the direction where the
last new idea came from.
This is parametrized by
the probability $p$ ($p\le 0.25$) to listen to one of the other 3 directions.

The fashions
in the system come from cultural centers.
Contrary to our previous model where only one cultural stronghold
is placed in the system {\it a priori}~\cite{lizana2011},
we here assume that a new cultural center can emerge
at any site with a small probability
$\pin$.
Such a site is recognized as a cultural
center as long as it has its own fashion and not invaded by
fashions from other sites. An existing cultural center
in addition broadcasts itself repeatedly by initiating
a new fashion with a rate $\pr$.

We perform Monte-Carlo simulation of the model with
parallel (synchronous) update of all agents.
In the model, a fashion $I(s,a)$ is characterized by its center $s$
(the site at which the idea started) and its age $a$ (how long time
ago the fashion originated at the center). At any time $t$ each site
$i$ have its current fashion $I_i(t)=I(s_i,a_i)$ and
its preferred direction $d_i(t)$,
from which this fashion was obtained.

The time step from
time $t$ to $t+1$ consists of the following procedures:
\begin{itemize}
\item[(i)]
{\sl Emergence of new cultural centers.}
With a probability $\pin$,
a site $i$ is randomly chosen out of all $N$ sites in the system
to become a new cultural center.
It starts its own new fashion, i.e., $I_i(t)$ is set to $I(i,0)$.
\item[(ii)]
{\sl Repeated broadcast by existing cultural centers.}
Each cultural center $i$ (i.e., $s_i=i$)
will start to spread a new fashion with probability $\pr$,
namely $I_i(t)$ is set to be $I(i,0)$.
Putting it differently, every cultural center can re-broadcast the same fashion
as a new one making it more appealing.
\item[(iii)]
{\sl Spreading of ideas.}
For each site $i$ in the system, the preferred site $d_i(t)$
is chosen with probability $1-3p$ ($p \leq 0.25$), or alternately
one chooses one of the other neighbor sites with the probability $p$.
The age $a_k$ of fashion $I_k(t)=I(s_k,a_k)$ at the chosen site $k$ is
compared with the age $a_i$ of the fashion at the site $i$.
If $a_k<a_i$, the site $i$ accepts the
fashion from the site $k$, namely set $I_i(t+1)=I_k(t)=I(s_k,a_k)$
and update its preferred direction to $d_i(t+1)=k$.
Otherwise the site $i$ keep its fashion unchanged, i.e.
$I_i(t+1)=I_i(t)=I(s_i,a_i)$ and keep its preferred direction
$d_i(t+1)=d_i(t)$.
If a site $i$ was a cultural center, the acceptance of competing idea
destroys its ability as a cultural center, hence it
stops repeatedly broadcasting new fashions.
\item[(iv)]
{\sl Update of time.}
The ages of all fashions on all sites are increased by one.
\end{itemize}

We simulate the model on an $L\times L$ square lattice
under the periodic boundary conditions.
We
will also consider the model on a map of Europe,
where the closed boundary conditions are imposed
toward the sea regions.
Initial condition is set so that
no one has opinion nor preferred directions (all the sites
weighted equally).
We only investigate properties of the system
after the number of cultural centers have reached
the steady state value.

\section{Results \label{sec:results}}
\subsubsection{Dynamics}
Dynamics of the model is depicted in
Fig.~\ref{fig:timeseries}, which presents time evolution on the European map.
In the top panel the respective cultures
are presented
with different colors.
The middle panel show current
distance of each sites from its respective cultural center,
defined as the place where its current fashion
was introduced.
The consecutive images
illustrate the dynamics of the system,
with meanderings
of borders, as well as emergence of
new cultural centers and disappearance of others.
Bottom panel presents the information pathways (river landscape),
based on the preferred direction for each site $d_i(t)$.
These arrows define the path to the cultural center to which every agent belongs.
Intensity of points in the river landscape indicates how many times information
was transmitted through every node, i.e. every time the idea is copied
the number of transitions
on all the nodes on the path (up to the origin) are increased by one.
It shows clear river basin structure centered around respective cultural centers,
much like what was obtained for the word spreading model of Ref.~\cite{lizana2011}.

\begin{figure}[!h]
 \centering
 \includegraphics[width=0.95\columnwidth]{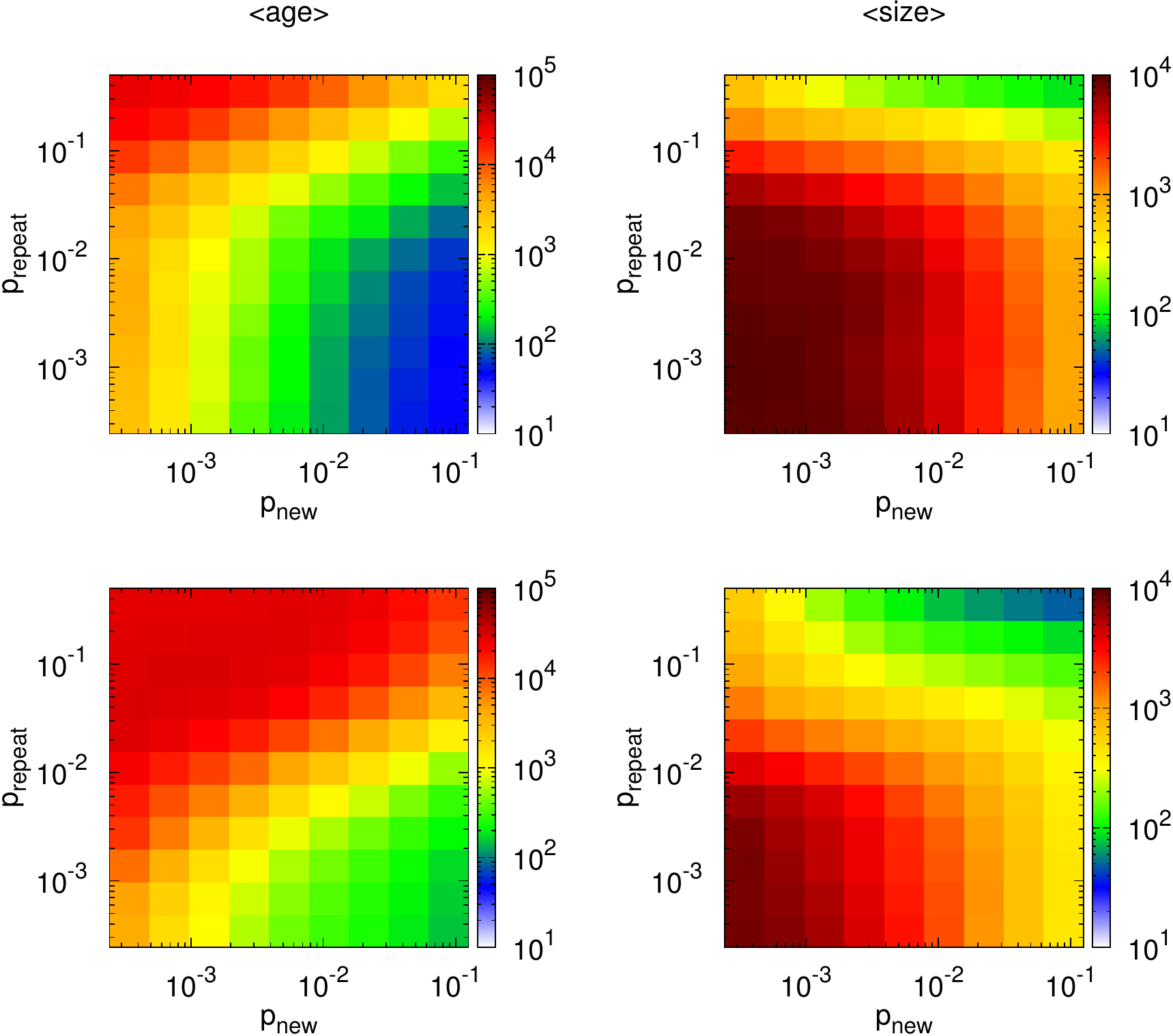}
 \caption{(Color on-line) Phase diagram showing average age, $\langle \mathrm{age} \rangle$, (left column) and average size, $\langle \mathrm{size} \rangle$, (right column) of cultural centers as the function of the probability of introduction of new cultures $\pin$ and the probability of resending new signals/rumor $\pr$. The system size is $100 \times 100$. The top row corresponds to $p=0.25$, while the bottom presents results for $p=0.01$.}
 \label{fig:phasediagram}
\end{figure}

\subsubsection{Analysis of parameters}
The role of the three parameters $p$, $\pin$, and
$\pr$ is summarized in Fig.~\ref{fig:phasediagram},
representing behavior of
$100 \times 100$ system with periodic boundary conditions.
In Fig.~\ref{fig:phasediagram},
the average age, $\langle \mathrm{age} \rangle$, (left panel)
and average size, $\langle \mathrm{size} \rangle$, (right panel) of cultural centers
are presented as
a function of $\pin$ and $\pr$,
with $p=0.25$ (top panel) and $p=0.01$ (bottom panel) respectively.
Note that $p=0.25$ examines the
case where
there is no conservativeness in the dynamics.
As $p$ becomes smaller, direction to existing cultural centers are preferred
and it becomes difficult for a new cultural center to emerge,
as can be seen in the longer lived cultural centers for smaller $p$.

For a given $p$, cultures live longer for smaller $\pin$ and larger $\pr$,
because small $\pin$ decreases emergence of new cultural centers,
while large $\pr$ assures stability of the cultural center, i.e. constant
broadcasting of new signals is vital for maintenance of cultural centers.
In the opposite limit of large $\pin$ and small $\pr$ the ongoing strong
competition between various ideas results in short living cultural centers.

The biggest cultural centers are observed for small $\pin$ and small $\pr$,
see the right column of Fig.~\ref{fig:phasediagram}.
Such a combination of parameters reduces the competition among
cultural centers and guarantees that each culture has enough time to spread
over the whole system, resulting in one dominating culture of the system size.
The opposite limit of large $\pin$ and large $\pr$,
on the other hand, means frequent emergence of new cultural centers
which survives relatively well,  resulting in the coexistence of many
small cultural centers.

We checked the effect of the system size by comparing the results with
$200 \times 200$ system,
and confirmed that the data in Fig.~\ref{fig:phasediagram}
collapses onto the data from the bigger system
very well as long as the data with the same
$p, \pr,$ and $\pin/N$ are compared.
The only exceptions are observed when the average size of fashion reaches the system size (data not shown).

\subsubsection{Competition between cultural centers}
\begin{figure}[!h]
 \centering
 \includegraphics[width=0.95\columnwidth]{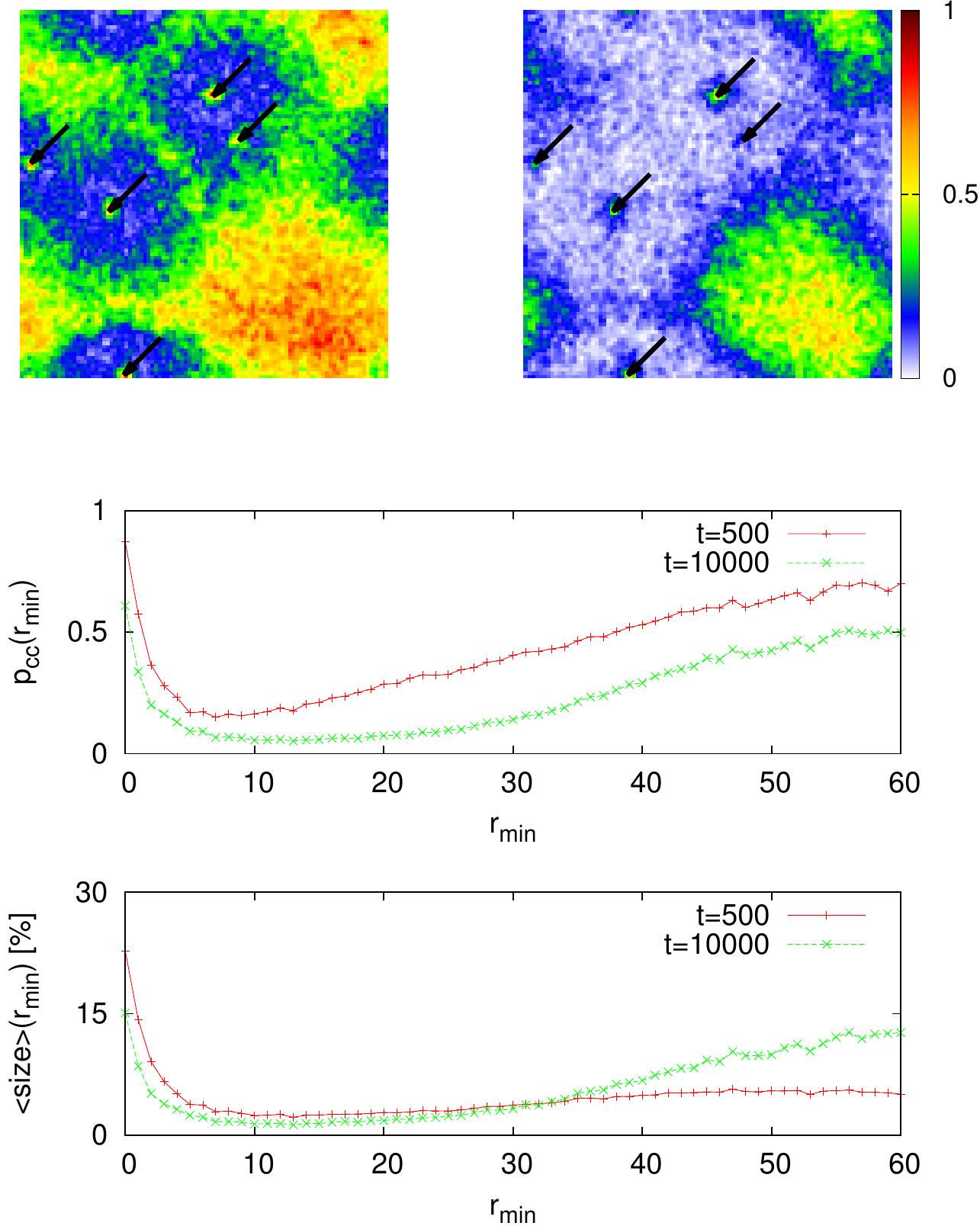}
 \caption{(Color on-line) Probability to survive as cultural center, $\pcc$, (middle row) and average size, $\langle \mathrm{size} \rangle$, (bottom row) as a function of the distance to the nearest cultural center, $r_{\mathrm{min}}$. Various lines correspond to various times $t=500$ and $t=10~000$. The average size is presented as the ratio of the whole system size. The top row present probability to survive as a cultural center as a function of the position $(x,y)$ at $t=500$ (left map) and $t=10~000$ (right map). The arrows in maps indicate positions of cultural centers in initial configuration. Simulation parameters: $p=0.01$, $\pin=2^{-10}$ and $\pr=2^{-6}$. The system size is $100 \times 100$.}
 \label{fig:ccsurvival}
\end{figure}

The interesting aspect of our model is the ongoing replacement of old cultural centers with new ones,
a dynamics primarily governed by
the emergence of new cultural centers.
To quantify this, we examine where
new cultural centers tend to emerge, when there are already
established cultural centers in the system.
 We re-run history multiple times with using a given snapshot of a $100 \times 100$ system
as an initial configuration.
For this initial configuration, we
insert a new cultural center to a site at $(x,y)$ at time zero, and run the simulation to see
how long the new center survives, i.e. until it is overwritten by fashions from neighbor centers.
 This
procedure is repeated 25 times
for each site
to estimate the survival probability
as a function of time $t$.

The top panel of Fig.~\ref{fig:ccsurvival} presents the 2D map of the survival probability
after 500 time steps (left map) and 10~000 time steps (right panel) as a function of the
position $(x,y)$ where new cultural center was inserted.
In the plot, the positions of the cultural centers that
exist in the initial configuration are marked by arrows.
We see that new cultural centers are successful either when
emerging very close to the existing cultural centers
or when exploring remote regions.

In order to see this tendency more clearly,
the middle panel presents survival probability, $\pcc$, as a function of the
distance to the closest cultural center, $\rmin$.
Similarly, the bottom panel of Fig.~\ref{fig:ccsurvival} shows the average size of the culture as
a function of $\rmin$.
We can see that both the survival probability and average size are a non-monotonous function
of the distance to the closest cultural center.
The insertion points located very close to the existing cultural
centers lead to maximal chances of surviving ($\pcc$) and
largest size of cultures ($\langle \mbox{size} \rangle$).
Both $\pcc$ and $\langle \mbox{size} \rangle$
drop quickly with distance to the existing cultural center $\rmin$
and then show a slow recovery with the distance $\rmin$.
Namely, new centers either explore a strategy of acquiring  the
existing network by taking over a previous center, or have to
explore the weaknesses of boundary regions to build its own network of influence.
The difficulty in building, rather than taking over an empire, is also reflected to the smaller size of new
cultures emerging in distant regions, compared to new cultures build on deposing an existing ruler-ships.
For larger times, both survival probability ($\pcc$) and
 average size ($\langle \mathrm{size} \rangle$) decrease for all the places
 (Fig.~\ref{fig:ccsurvival} middle and bottom panels)
 since competition with existing cultural centers makes for a finite extinction rate.

\subsubsection{Analysis of one dimensional model}\label{onedim}

\begin{figure}
\includegraphics[width=0.95\columnwidth]{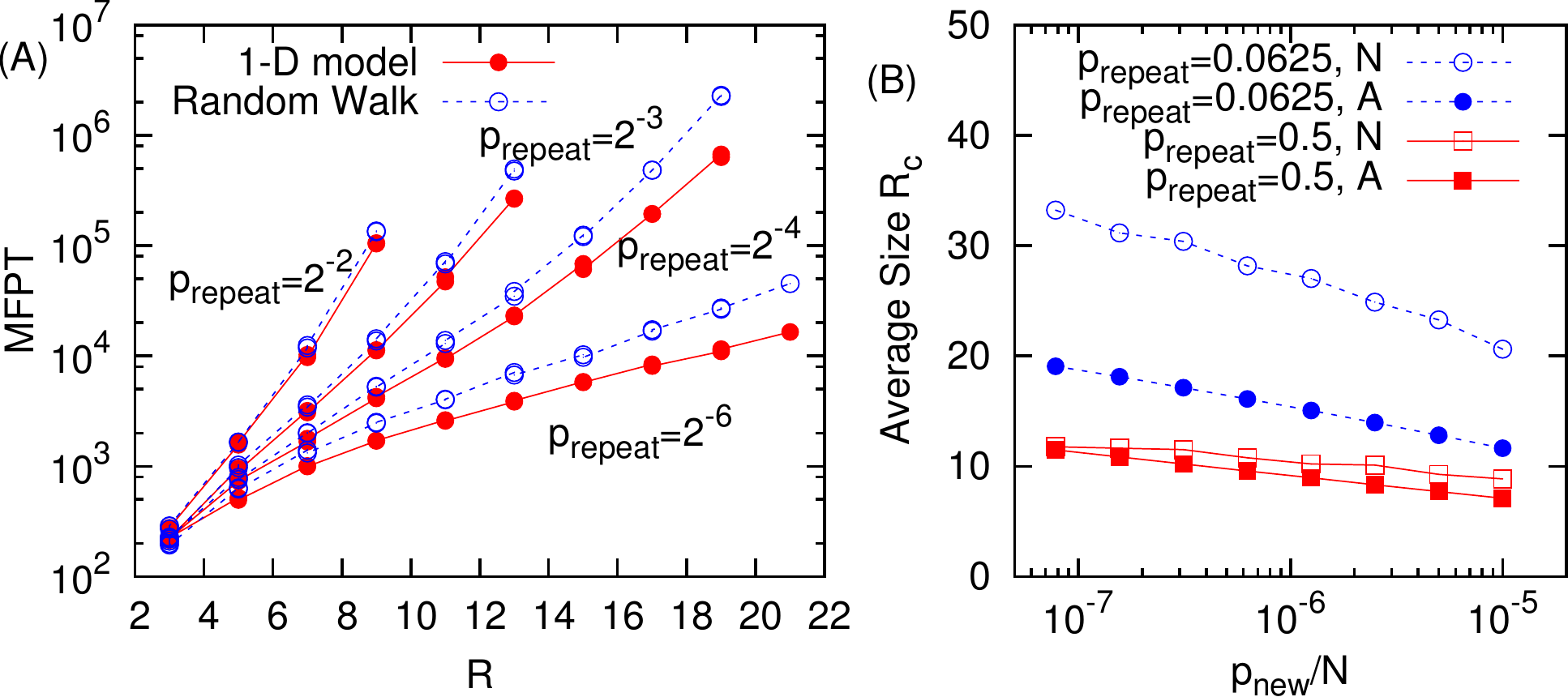}
\caption{(Color on-line) (A) Mean first passage time (average survival time of
a cultural center) vs distance
between two cultural centers $R$ in
one dimensional model (filled circles)
and simplified random walk model (open circles),
with $p=0.01$. One can see good agreement for
$p \le \pr$. The deviation grows for
$p \ge \pr$, but the qualitative behavior is captured
in the random walk model.
(B) Average size of cultural area in one dimension.
``N'' denote the numerical simulation and ``A'' denote the analytical
estimate. As $\pr$ becomes smaller the disagreement between
the simulation and the analysis becomes larger.
}\label{MFPT}
\end{figure}

The reported features of the present
culture spreading model can be understood qualitatively by
considering a simplified one dimensional
model of the random walk of the boundary between
two cultural centers.

Suppose that there is a cultural center $C_0$ at the site 0
and another cultural center $C_1$ at the site $R$.
We consider the motion of the left most site $r$ that belongs to $C_0$.
Moreover, we assume that no additional cultural center appears, i.e.  $\pin=0$.
In this limit, the age $a_r$ of the fashion at site $r$
which belongs to $C_0$,
and the age $a_{r+1}$ of the fashion at site $r'=r+1$,  which belongs to $C_1$,
can be approximated as
\begin{equation}
a_r\approx \frac{r}{1-p}+\tau,\quad
a_{r'}\approx \frac{R-r'}{1-p}+\tau',
\label{age}
\end{equation}
respectively.
In Eq.~(\ref{age}), $\tau$ and $\tau'$ are independent discrete
stochastic variables both having a probability distribution
\begin{equation}
P(\tau)=\pr (1-\pr)^\tau.
\label{taudist}
\end{equation}
For the sake of simplicity, the stochasticity of the fashion propagation
in the preferred direction and the time correlation of the
age are ignored, see Eq.~(\ref{age}).

We can calculate the
rate that the position $r$ of the left most site that
belongs to $C_0$ decreases (increases) by one,
which happens if $a_r>a_{r+1}$ ($a_r<a_{r+1}$) when
the site $r$ ($r+1$) listen to site $r+1$ ($r$)
with probability $p$.
The explicit form of the rates are given in the Appendix.
Recall that the presented derivation is valid when $p\ll \pr$, where
the approximation of the age in Eq.~(\ref{age}) is justified.

In the left panel of Fig.~\ref{MFPT}, we compare the mean first passage
time (MFPT) of the boundary starting at $r=(R-1)/2$
to reach one of the cultural centers,
$T(p,\pr,R)$,
which gives the typical survival time
of established cultural centers separated by distance $R$.
The simplified random walk model of the boundary between two cultural centers
agree reasonably well with the full one dimensional simulation results.
We also checked that the level of agreement improves when $p\ll \pr$
(data not shown).
We can see that the survival time is
longer for larger $\pr$ and increase exponentially for large
$R$.

It is more intuitive to interpret these results by
making continuous approximation for space and
time and deriving the Fokker-Planck equation
for the probability $P(x,t)$ that the boundary is
at the position $x=r-(R-1)/2$ (therefore the cultural centers are
located at $x\approx \pm R/2$)
at time $t$. The resulting equations are (see the Appendix),
\begin{equation}
\frac{\partial P(x,t)}{\partial t}=-\frac{\partial}{\partial x} J(x,t),
\label{FP}
\end{equation}
where
\begin{equation}
J(x,t)=-p\frac{dU(x)}{dx}P(x,t)-\frac{\partial}{\partial x}\left[D(x)P(x,t)\right],
\end{equation}
and
\begin{equation}
U(x)=|x|-\ell \left[ 1-\exp\left(-\frac{|x|}{\ell}\right) \right],
\label{eq:potential}
\end{equation}
\begin{equation}
D(x)=\frac{p}{2}\left[1- \frac{\pr}{1-\pr}\exp\left(-\frac{|x|}{\ell}\right)\right],
\label{diffconst}\\
\end{equation}
\begin{equation}
\ell(p,\pr)= \left[-\frac{2\ln(1-\pr)}{1-p}\right]^{-1}
\approx \frac{1-p}{2\pr}.
\label{ell}
\end{equation}
The potential $U(x)$ has a minimum at $x=0$ with harmonic behavior
($U(x)\approx x^2/(2\ell (p,\pr))$ for $|x| \ll \ell (p,\pr)$)
while $U(x)$ grows linearly with $|x|$ for large values of argument ($|x| \gg \ell (p,\pr)$).
On the other hand, the dependence of the diffusion coefficient
$D(x)$ on position is rather weak and it can be considered as $D\approx p/2$.
The probability $p$ also defines the mobility, therefore the timescale of the
random walk of the boundary is proportional to $1/p$.

Now we can estimate the typical size of the
cultural area or the typical distance between centers $R_c$.
Suppose there is a cultural center at
position $0$, and a new cultural center is inserted
at a distance $R$. If new cultural center will not be
inserted, the time scale when
one of them will be overwritten by the other one can be
estimated by the mean first passage time, $T(p,\pr,R)$,
starting from the stable point $x=0$.
During this period,
however, a new cultural center can be inserted between
$0$ and $R$ with a probability $\approx \pin R/N$, where
$\pin /N$ is the insertion probability per site.
Therefore, the insertion and the coarsening
balance leads to
\begin{equation}
T(p,\pr,R_c)/2 = N/(\pin R_c).
\label{estimate}
\end{equation}
The factor 1/2 on left hand side of Eq.~(\ref{estimate}) comes from the fact that
each centers are competing with two other centers
on both sides. The comparison of the average cultural
center size from the one dimensional  simulation
and the estimate based on Eq.~(\ref{estimate})
with the mean first passage time $T$ evaluated under continuum approximation,
see Eq.~(\ref{MFPTeq}), are shown in
the right panel of Fig.~\ref{MFPT}.
The agreement is satisfactory for large $\pr$ but
becomes worse for smaller $\pr$. One of the reason
for disagreement is that $\ell(p,\pr)$ becomes considerably
large for small $\pr$. Consequently, the potential becomes flatter
and newly inserted cultural centers have higher probability
to be overwritten before the boundary reaches the central point ($x=0$),
which enhances coarsening hence increasing the average size
of cultural area.

In the two dimensional case, the size becomes proportional to
$R_c^2$, but other parameter dependence are expected to be
qualitatively the same.
The mean first passage time $T$ is proportional to $1/p$, see Eq.~(\ref{MFPTeq}),
which is the time scale of the dynamics,
while rapidly growing function of both
$\pr$ and $R$. Therefore, the average size $R_c^2$
estimated with (\ref{estimate}) are expected to
decrease with $p$, $\pr$, and $\pin$, what is
consistent with Fig.~\ref{fig:phasediagram}.

\subsubsection{Replaying history of Europe\label{sub:rep}}
\begin{figure}[!h]
 \centering
 \includegraphics[width=0.7\columnwidth]{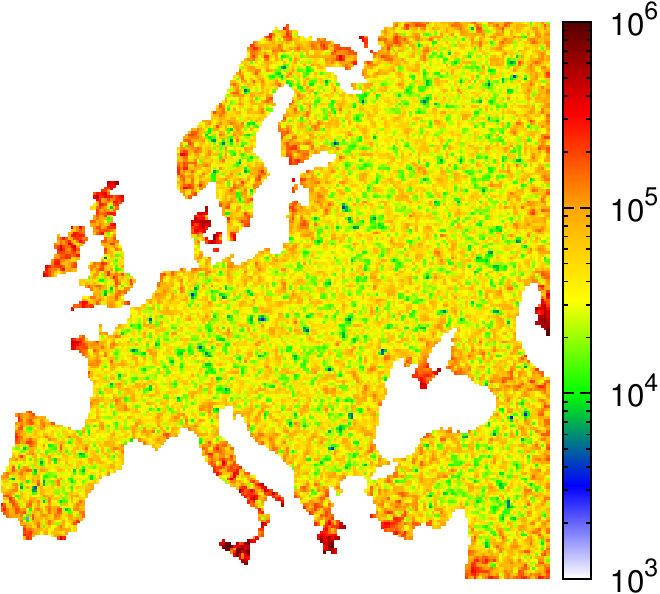}
 \caption{(Color on-line) The frequency histogram, presenting the fraction of time for which each point constituted a cultural center. Parameters: $\pin=0.004$ and $\pr=0.015$, $p=0.01$. Simulation time $T=10^8$ Monte Carlo steps.}
 \label{fig:ccfrequency}
\end{figure}
Finally,
we examine the dynamics of our model
on the European map on a $200\times 200$ square lattice
with a land mass that consists of 24~156 sites.
The only constraint that the map give is its boundary conditions
where in particular the sea is impenetrable, and thus information cannot travel across seas.
We use parameters where roughly 20 cultural centers
coexists (with $p=0.01$, $\pin=0.004$ and $\pr=0.015$;
as also used in Fig. 1).
Assuming that the fastest time it takes a rumor to cross Europe is about 10 years,
the corresponding number of updates on our 200$\times200$ will be $\sim 200$.
In this perspective the subsequent snapshots in Fig. 1 correspond to 50 years,
a timescale where changes on the European map indeed occurred throughout the last millennium.
Fig.~\ref{fig:ccfrequency} shows a fraction of time for which respective sites
was a cultural center, illustrating that
centers tend to be more stable or more often established at the tips of peninsulas or other
remote regions of the continents.
That is, the chance to be invaded by competing fashions and cultures diminishes
in these remote edge regions.

Additionally, we have checked robustness
of the observed patterns in Fig.~\ref{fig:ccfrequency}.
More precisely, we have constructed frequency histograms for size
dependent broadcast probability $\pr$  ($\pr \sim \mbox{size}/\mbox{system size}$).
This modification is based on a picture that
new fashions come more frequently when the size of the
culture is bigger, which in consequence
makes a bigger culture more stable and convincing.
The presence of such a positive feedback
weakened the contrast between peripheries and internal regions.
Nevertheless, distant points remained harder to be invaded
than internal points (data not shown).
We also studied the effect of short-cuts that connect
two randomly chosen remote sites,
having the possibility to building main roads between
cities in mind. Presence of short-cuts further reduced the contrast between remote and central points
(data not shown).

It should also be pointed out that the studied model does not account
for geographical constraints like rivers, mountain chains, climate and population distribution,
which are crucial for the spread of fashions
and cultures in real life situations.
It is assumed that transmission of fashions are purely local
and in particular that fashions does not travel overseas.
As a consequence, the European simulation is more an illustration of the
basic principle of the model than a valid simulation of available
information highways on an ancient European landscape.

The incorporation of mentioned constraints can significantly
change the properties of the model.
Rivers and roads constituted information paths in pre-telegraph Europe while mountain chains provide natural communication barriers.
Contrary to geographical landscape, the role of varying population density is more complex and less apparent.
 On the one hand, it is natural to imagine that large population density
leads to larger creativity to start a new culture, and
the number of people sharing the same culture also affects
the ability of the culture to convince other people.
On the other hand,  it is likely that there is a positive feedback from cultural center
to the local population,  i.e.,
 larger population density appears in places which are close to existing cultural centers.
It would be an interesting future project to incorporate such an effect in the present model.

\section{Summary and discussions\label{sec:summary}}
We have explored a simple model for emergence and decline of cultural strongholds,
parametrized with rigidity in local social network
$p$, probability of emergence $\pin$ and the probability
at which an existing cultural center broadcast fashions $\pr$.

The overall assumption of the model was the postulate that individuals
always accept the newest (locally) available viewpoint as their own,
but obviously cannot adopt a viewpoint that is not available
in their social surroundings.
Social surroundings were
restricted to their 4 nearest neighbors on a 2D square lattice,
and further biased with their preference for listening in the direction where
they last obtained a useful information.
As probability to listen in other competing directions decreases ($p\rightarrow 0$)
the cultural map freezes into many small regions.

The acceptance of a fashion according to its
age only is an important feature of our model.
In some models of opinion dynamics, on the other hand,
a set of rules, which on average makes
an agent accept a fashion shared by majority,
have been adopted (e.g., voter models \cite{holley1975}
or majority rules models \cite{galam2002}).
Incorporation of such rules to the present model
should make it more difficult for a new culture
center to emerge and grow,
though quantitative effect depends on the exact rules.

It is also
worth mentioning the relation between the present model
and Axelrod model \cite{axelrod1997},
which is a widely accepted model of formation of cultural area
both by social scientists and physicists \cite{CastellanoReview2009}.
In the Axelrod model, each agents has a set of opinions as a vector
and interact with the neighboring agents according to the
overlap of opinions: It is more likely to interact when
the opinions are close, and when they interact
the agent copies one of the different opinion from the neighbor
to its opinion set. In this dynamics, the
conservativeness is taken into account as
tendency to talk to the agents that has close opinion,
and the cultural area is formed as the agents align their
opinions with their neighbors.
In a sense, a culture spontaneously appears via interactions between
agents in this model.
The model can show coexistence of multiple cultural areas
but it turned out that coexistence is unstable against
spontaneous flipping of the opinions \cite{Klemm2003},
and several modifications of the model has been and
is studied to realize stable coexistence of cultures
 \cite{CastellanoReview2009}.

On the contrary, in our model cultural area is
defined as the area that shares the same information source.
The random appearance of the new cultural center, which
can be viewed as a spontaneous change of the opinion set in
Axelrod model, is actually the important feature to
keep the multiple cultural centers against one culture
taking over the whole system.
The key feature of our model to make this possible
is the importance of newer information,
which give some chance for newcomer to win against
existing cultural centers.
It can be interesting to add a similar feature to
the Axelrod model, i.e.,
give some rate to renew opinions and value newer information more to
verify whether multiple cultures can coexist in that case.


\begin{acknowledgments}
The authors acknowledge the Danish National Research Foundation for financial support.
Computer simulations have been performed at the Academic
Computer Center, Cyfronet AGH (Krak\'ow, Poland) and CMOL Niels Bohr Institute.

\end{acknowledgments}

\appendix
\section{1D random walk model of the boundary between
two cultural centers}

Here we derive the simplified one dimensional
model of the random walk of the boundary between
two cultural centers.
As summarized in subsection \ref{onedim},
two cultural centers $C_0$ and $C_1$ are located at the site 0 and
at the site $R$, respectively.
We analyze
the motion of the left most site $r$ that
belongs to $C_0$ (the site $r+1$ belongs to $C_1$).

The rate the position $r$ decreases (increases) by one, $W_{r\to r-1}$
($W_{r\to r+1}$),
is given by the probability that $a_r>a_{r+1}$ ($a_r>a_{r+1}$)
when the site $r$ ($r+1$) listens to the unpreferred direction $r+1$ ($r$)
at a given time step. From (\ref{age}) and (\ref{taudist}) we get (note that $0<r<R$)
\begin{equation}
W_{r \to r-1}  =
\begin{cases}
p\frac{(1-\pr)^{\frac{R-2r-1}{1-p}+1}}{2-\pr} & \mbox{for}\;r\le \frac{R-p}{2}, \\
p\left[1-\frac{(1-\pr)^{-\frac{R-2r-1}{1-p}}}{2-\pr} \right] & \mbox{for}\;  r>\frac{R-p}{2},
\end{cases} \nonumber
\label{eq:rm}
\end{equation}
and
\begin{equation}
W_{r \to r+1}=
\begin{cases}
p\left[1-\frac{(1-\pr)^{\frac{R-2r-1}{1-p}}}{2-\pr}\right] & \mbox{for}\;\;  r\le \frac{R-p}{2},\\
p\frac{(1-\pr)^{-\frac{R-2r-1}{1-p}+1}}{2-\pr} & \mbox{for}\;  r>\frac{R-p}{2}.
\end{cases} \nonumber
\label{eq:rp}
\end{equation}
Using above transition rates
one can write the master equation for the
probability density $P_{r,t}$ that the boundary is at site $r$ at time $t$:
\begin{eqnarray}
P_{r,t+1}-P_{r,t}&=&W_{r-1\to r} P_{r-1,t}-W_{r \to r+1} P_{r,t} \nonumber \\ \nonumber
&&+W_{r+1\to r} P_{r+1,t}-W_{r\to r-1}P_{r,t}.
\end{eqnarray}
Assuming that the time step and the lattice spacing are small,
we obtain the following Fokker-Planck equation~\cite{VanKampenBook}
\begin{eqnarray}
\frac{\partial P(r,t)}{\partial t}
&=& \frac{\partial}{\partial r}\left[(W(r\to r-1)-W(r\to r+1))P(r,t)\right] \nonumber \\ \nonumber
& + &\frac{1}{2}\frac{\partial^2}{\partial r^2}\left[(W(r\to r-1)+W(r\to r+1))P(r,t)\right].
\end{eqnarray}
Substituting $r$ with $x=r-(R-1)/2$ one gets Eqs.~(\ref{FP}) to (\ref{ell}).

The mean first passage time $T$ starting
from $x=0$ to reach $x=\pm R/2$ is given by the closed formula~\cite{VanKampenBook}
\begin{equation}
T(p,\pr,R)=-\int_{-R/2}^0ds\left[e^{\Phi(s)}
\int_0^s \frac{e^{-\Phi(y)}}{D(y)}dy
\right], \nonumber
\end{equation}
where
\begin{equation}
\Phi(x)=\int_0^x \frac{p U'(y) }{D(y)}dy. \nonumber
\end{equation}
Approximating  $D(x)$ with $p/2$, see Eq.~(\ref{diffconst}), one gets
\begin{equation}
T(p,\pr,R)\approx -\frac{2}{p}\int_{-R/2}^0ds\left[e^{2U(s)}
\int_0^s {e^{-2U(y)}}dy
\right], \nonumber
\label{MFPTeq}
\end{equation}
from which Eq.~(\ref{estimate}) can be evaluated
using Wolfram Mathematica.



\begin{thebibliography}{26}
\expandafter\ifx\csname natexlab\endcsname\relax\def\natexlab#1{#1}\fi
\expandafter\ifx\csname bibnamefont\endcsname\relax
 \def\bibnamefont#1{#1}\fi
\expandafter\ifx\csname bibfnamefont\endcsname\relax
 \def\bibfnamefont#1{#1}\fi
\expandafter\ifx\csname citenamefont\endcsname\relax
 \def\citenamefont#1{#1}\fi
\expandafter\ifx\csname url\endcsname\relax
 \def\url#1{\texttt{#1}}\fi
\expandafter\ifx\csname urlprefix\endcsname\relax\def\urlprefix{URL }\fi
\providecommand{\bibinfo}[2]{#2}
\providecommand{\eprint}[2][]{\url{#2}}


\bibitem{ackland2007}
G. Ackland, M. Signitzer, K. Stratford, and M. Cohen, Proceedings of the
  National Academy of Sciences {\bf 104},  8714  (2007).

%
%
%


\bibitem[{\citenamefont{Axelrod}(1997)}]{axelrod1997}
\bibinfo{author}{\bibfnamefont{R.}~\bibnamefont{Axelrod}},
 \bibinfo{journal}{J. Conflict Resolt.} \textbf{\bibinfo{volume}{41}},
 \bibinfo{pages}{203} (\bibinfo{year}{1997}).

\bibitem[{\citenamefont{Castellano et al.}(2009)}]{CastellanoReview2009}
\bibinfo{author}{\bibfnamefont{C.}~\bibnamefont{Castellano}},
\bibinfo{author}{\bibfnamefont{S.}~\bibnamefont{Fortunano}},
\bibnamefont{and}
\bibinfo{author}{\bibfnamefont{V.}~\bibnamefont{Loreto}},
 \bibinfo{journal}{Rev. Mod. Phys.} \textbf{\bibinfo{volume}{81}},
 \bibinfo{pages}{591} (\bibinfo{year}{2009}).



\bibitem{holley1975}
R.~A. Holley and T.~M. Liggett, Ann. Probab. {\bf 3}, 643 (1975).

\bibitem{lewenstein1992}
M. Lewenstein, A. Nowak, and B. Latan\'e, Phys. Rev. A {\bf 45}, 763 (1992).

\bibitem{galam2002}
S. Galam, Eur. Phys. J. B {\bf 25}, 403 (2002).

\bibitem{sznajd-weron2000}
K. Sznajd-Weron and J. Sznajd, Int. J. Mod. Phys. C {\bf 11}, 1157 (2000).

\bibitem{deffuant2000}
G. Deffuant, D. Neau, F. Amblard, and G. Weisbuch, Adv. Complex Syst. {\bf 3},
 87 (2000).

\bibitem{hegselmann2002}
  R.~Hegselmann, and U.~Krause,
  J. Artif. Soc. Soc. Simulat. \textbf{5}(3) (2002).


\bibitem{pineda2011}
M. Pineda, R. Toral, and E. Hern{\'a}ndez-Garc{\'\i}a,
  Eur. Phys. J. D \textbf{62}, 109 (2011).

\bibitem{huet2008}
S. Huet, G. Deffuant, and W. Jager,
Adv. Complex Syst. \textbf{11}, 529 (2008).



\bibitem{rosvall2009}
M. Rosvall and K. Sneppen, Phys. Rev. E {\bf 79}, 026111 (2009).



\bibitem[{\citenamefont{Stiglitz and Weiss}(1981)}]{stiglitz1981credit}
\bibinfo{author}{\bibfnamefont{J.}~\bibnamefont{Stiglitz}} \bibnamefont{and}
 \bibinfo{author}{\bibfnamefont{A.}~\bibnamefont{Weiss}},
 \bibinfo{journal}{Am. Econ. Rev.}
 \textbf{\bibinfo{volume}{71}}, \bibinfo{pages}{393} (\bibinfo{year}{1981}).

\bibitem[{\citenamefont{Rosvall and Sneppen}(2006)}]{rosvall2006self}
\bibinfo{author}{\bibfnamefont{M.}~\bibnamefont{Rosvall}} \bibnamefont{and}
 \bibinfo{author}{\bibfnamefont{K.}~\bibnamefont{Sneppen}},
 \bibinfo{journal}{EPL} \textbf{\bibinfo{volume}{74}},
 \bibinfo{pages}{1109} (\bibinfo{year}{2006}).

\bibitem[{\citenamefont{Lizana et~al.}(2010)\citenamefont{Lizana, Rosvall, and
 Sneppen}}]{lizana2010time}
\bibinfo{author}{\bibfnamefont{L.}~\bibnamefont{Lizana}},
 \bibinfo{author}{\bibfnamefont{M.}~\bibnamefont{Rosvall}}, \bibnamefont{and}
 \bibinfo{author}{\bibfnamefont{K.}~\bibnamefont{Sneppen}},
 \bibinfo{journal}{Phys. Rev. Lett.} \textbf{\bibinfo{volume}{104}},
 \bibinfo{pages}{040603} (\bibinfo{year}{2010}).





\bibitem[{\citenamefont{Yanagita}(1930)}]{Yanagita}
\bibinfo{author}{\bibfnamefont{K.}~\bibnamefont{Yanagita}},
 \emph{\bibinfo{title}{{Kagyuko}}} (\bibinfo{publisher}{Toko Shoin},
 \bibinfo{year}{1930}).

\bibitem{ramsey1982language}
S. R. Ramsey, J. Jpn. Stud. {\bf 8}, 97 (1982).


\bibitem[{\citenamefont{Lizana et al.}(2011)}]{lizana2011}
\bibinfo{author}{\bibfnamefont{L.}~\bibnamefont{Lizana}},
\bibinfo{author}{\bibfnamefont{N.}~\bibnamefont{Mitarai}},
\bibinfo{author}{\bibfnamefont{K.}~\bibnamefont{Sneppen}},
\bibinfo{author}{\bibfnamefont{H.}~\bibnamefont{Nakanishi}},
 \bibinfo{journal}{Phys. Rev. E} \textbf{\bibinfo{volume}{83}},
 \bibinfo{pages}{066116} (\bibinfo{year}{2011}).





\bibitem{Klemm2003}
K.~Klemm, V.~M.~Eguiluz, R.~Toral, and M.~San~Miguel,
Phys. Rev. E {\bf 67}, 045101(R) (2003).


\bibitem{VanKampenBook}
N.~G.~Van Kampen, {\sl Stochastic processes in
physics and chemistry} (Elsevier, Amsterdam 1997).




\end{thebibliography}
\end{document}